# Design of an assistive trunk exoskeleton based on multibody dynamic modelling

*Pierre Lifeng Li[1,2,3], Sofiane Achiche[1,2], Laurent Blanchet[1,2,3], Samuel Lecours[1,2,3], Maxime Raison[1,3]

[1] Department of mechanical engineering, École Polytechnique de Montréal, 2900 Boul. Édouard-Montpetit, Montreal (Qc), Canada H3T 1J4
[2] Laboratoire de Conception de Systèmes Intelligents et Mécatroniques (CoSIM), École Polytechnique de Montréal, 2900 Boul. Édouard-Montpetit, Montreal (Qc), Canada H3T 1J4
[3] Rehabilitation Engineering Chair Applied in Pediatrics (RECAP), Marie Enfant Rehabilitation Centre, 5200 Bélanger, Montreal, Quebec, Canada H1T1C9

***Correspondence Address:** Pierre Lifeng Li
Laboratoire de Conception de Systèmes Intelligents et Mécatroniques (CoSIM), École Polytechnique de Montréal, 2900 Boul. Édouard-Montpetit, Montreal (Qc), Canada H3T 1J4
Email address: pierre-lifeng.li@polymtl.ca

## Abstract

Low back pain is one of the most common musculoskeletal disorder. To reduce its incidences, some back exoskeletons have been designed and already commercialized. However, there is a gap between the phases of device and testing on subjects. In fact, the main unsolved problem is the lack of realistic simulation of human-device interaction. Consequently, the objective of this paper is to design a 3D multibody model of the human body that includes the full thoracic and lumbar spine combined with a low back exoskeleton, enabling to analyze the interactions between them. The results highlight that the use of the exoskeleton reduces the net torque in the lower lumbar spine but creates normal forces transmitted through the thighs and the pelvis, which should also be considered for low back pain. As a perspective, this model would enable to redesign low back exoskeletons reducing both torques and forces in the human joints in a realistic dynamic context.

## 1. Introduction

Low back pain (LBP) is one of the most common musculoskeletal disorder (MSD), e.g. 26.9% of MSDs in the United States [1]. Workers in the United States industrial sector are the most likely to develop LBP, as the prevalence of disabling LBP represents 28%, i.e. over one fourth, of that population [3]. Industries reporting the highest risk and the most cases of LBP are, respectively, lumber and building material retailing, and construction [2]. According to Connen & al. [8], the cumulative low back load (CLBL), which is the equivalent torque produced by the low back while achieving a task, is a significant risk factor for LBP.

To reduce LBP prevalence and risk injury, wearable devices such as back belts and exoskeletons have been developed. These devices are designed to restrict harmful motions, increase human strength or distribute loads. Back exoskeletons typically decrease the load in the low back region and distributes it to the legs or directly to the ground. According to Ammendolia et al. [4], back belts have not shown conclusive evidence of preventing occupational LBP. Exoskeletons, however, have presented promising results [5]. These devices can be classified into two categories: passive (i.e. non-motorized, such as with spring-damper mechanisms) or active (i.e. motorized) exoskeletons. Some example of passive exoskeletons are Laevo [9], BNDR [11], PLAD [10] while Robomate [12] and Hyundai H-WEX [13] are examples of active devices.

To assess the back support effectiveness, experiments with human subjects were conducted by comparing results with and prior to the use of the prototypes. The results were extracted by different testing methods such as electromyographic (EMG) signals [10,11,13], musculoskeletal models [11,13] or qualitative questionnaires [9]. Some studies (e.g. [5]) also included the joint ranges of motion in their parameters and have indicated a reduction of these ones while wearing the device.

Simulations prior to functional experiments are critical to validate forces, torques and motions between the human and the exoskeleton [23] but are lacking in the literature. Multibody dynamic models have recently been used in different applications, such as optimal design of an elbow exoskeleton [24] and to determine custom sizing of a lower limb exoskeleton actuators [25]. A recent study [6] tried to overcome the problem by designing a 2D multibody model of a back support exoskeleton and a human. The human model was simplified to represent the entire back as two segments (lower and upper trunk). Another study [7], aiming to determine the influence of hinge positioning of their device, presented a 3D multibody human model with a back exoskeleton designed on the anterior side of the body. The human model trunk was simplified similarly as the previous mentioned study. These models [6, 7] present the spine as two segment thus simplifying spinal motion and load distribution. To evaluate the exoskeleton-human interaction more accurately, a more realistic model should be developed for exoskeleton-human interaction.

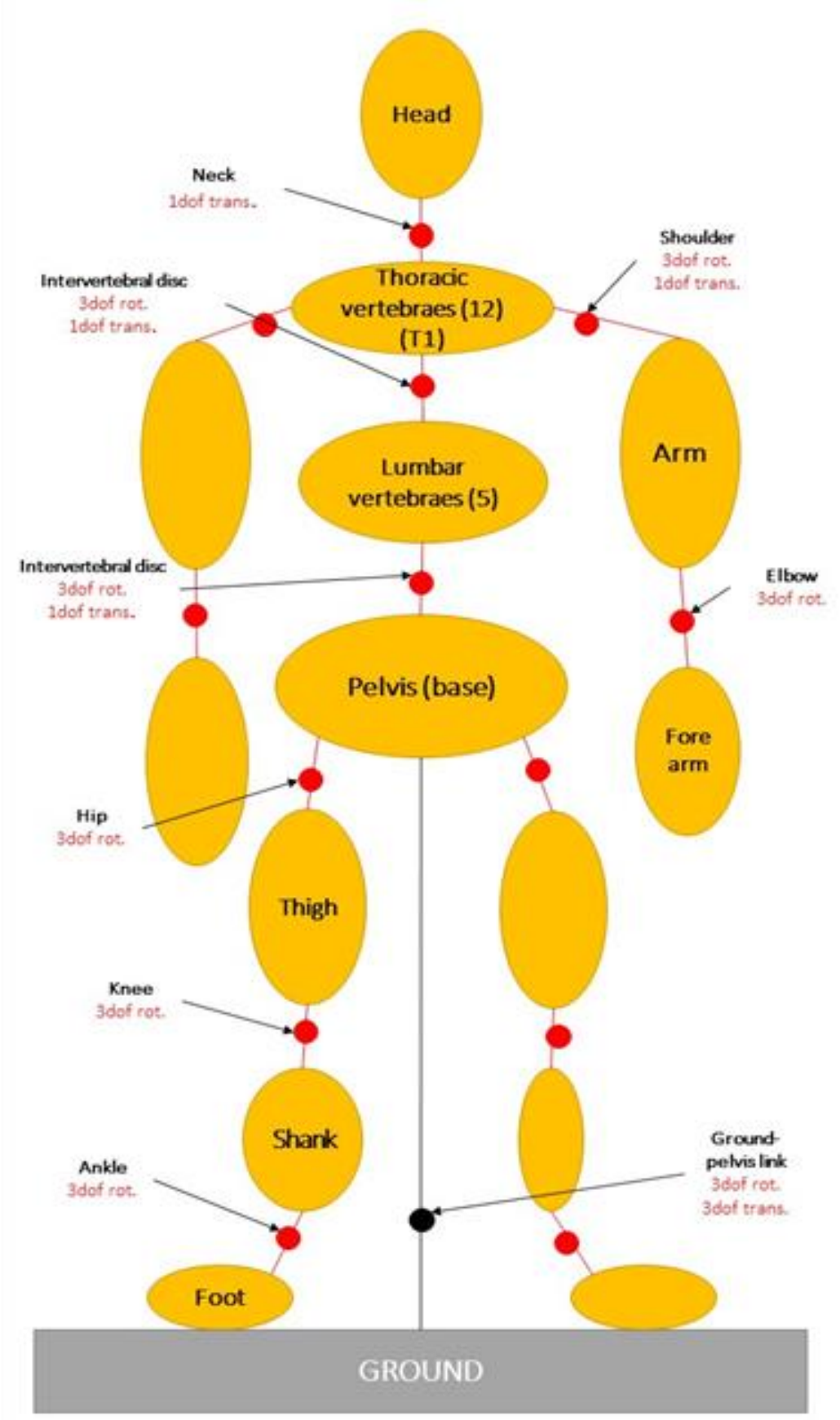

**Figure 1.** Human multibody model

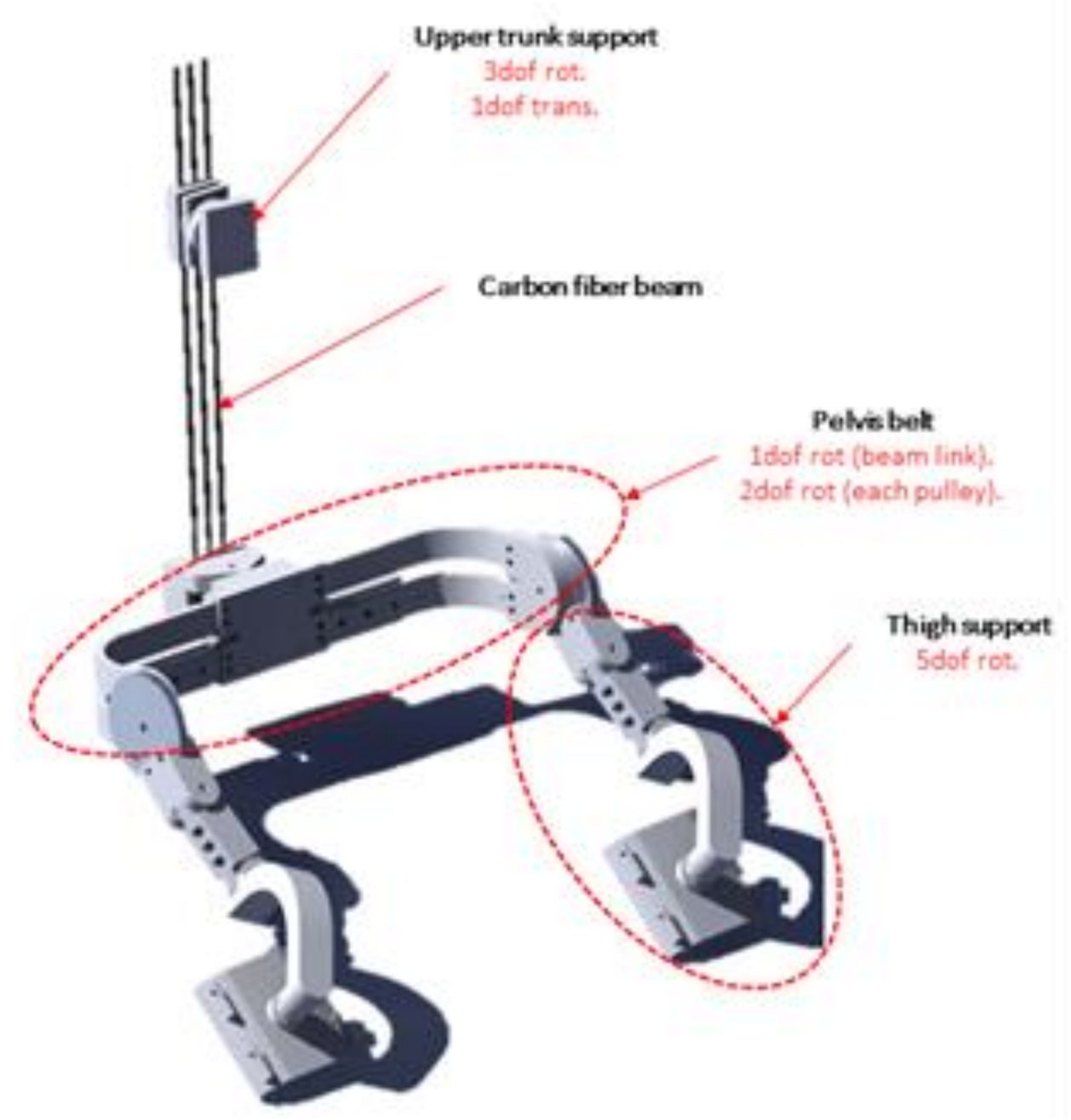

**Figure 2.** CATIA model of the exoskeleton

The aim of this study is to develop a novel 3D multibody human model which includes the complete thoracic and lumbar spine. Particularly, this study presents a passive (non-motorized) back exoskeleton with an articulated beam interacting with the human model. Performance is determined by parameters such as CLBL and net joint torques in other relevant joints and interfacing forces.

## 2. Methodology

### 2.1. Multibody model of the human body

The human model was implemented using ROBOTRAN software [20, 21], and is illustrated in Figure 1. The body geometric parameters and dynamic parameters (i.e. masses, centers of mass (CoM), moments of inertia) were based on a CATIA V5 manikin, representing a man of a total body mass of 68kg and a total body height of 1.70m. The model is composed of 29 rigid bodies: lower limb (6), pelvis (1), lumbar spine (5), thoracic spine (12), upper extremity (4) and head (1). As for each trunk slice, these parameters included the mass of the internal organs, while neglecting their relative movement.

The lower limbs were modeled as 3 bodies, i.e. thighs, shanks and feet, linked together by spherical joints and connected to the pelvis. The pelvis was considered the reference body and was linked to the base (ground) by translations and rotations in all 3 axis. The spine was modeled as a chain of 17 bodies linked with spherical joints and a translation in the longitudinal axis. The upper extremities (arms and forearms) were linked together by spherical joints and were connected to the first thoracic vertebrae of the spine. The head was also connected to the first thoracic vertebrae. The thoracic and lumbar vertebrae internal forces and moments were considered according to White & Panjabi's research [22]. Standing posture was assumed as neutral reference position, while angles are positive during flexion phase and negative during extension.

### 2.2. Exoskeleton Model

The exoskeleton was first modeled in CATIA V5 and personalized to the Manikin human model presented in section 2.1. The device was designed with four sub-systems (Figure 2):

1) Pelvis belt, interfacing with the posterior side of the waist. This piece contains 2 tension springs that connect to waist pulleys.
2) Upper trunk support, interfacing with the posterior side of the upper back.
3) 3 carbon fiber articulated beam (4mm x 4mm x 600mm) linking the pelvis belt and the upper trunk support.
4) Two thigh supports, interfacing with the user's legs and connected with the pelvis belt through pulleys at each side of the waist.

The material used to simulate the exoskeleton was plastic (except for the carbon fiber beams) giving the exoskeleton a total mass of 7.5kg.

This exoskeleton model was then implemented in ROBOTRAN along the human model (Figure 3). The thigh supports were composed of 4 rigid bodies (thigh support pulley, bridge, ball joint and interface). The thigh support pulley and the thigh support bridge were linked by using a revolute joint in the sagittal axis. The thigh support bridge and the thigh support ball joint were linked the same way. Between the thigh support ball joint and the thigh support interface, a ball cut was required to close the kinematic loop. At the interface between the thigh and the thigh support, a force driven prismatic joint was added to simulate the interfacing contact. The pelvis belt was modeled as one rigid body connected by two symmetrical revolute joints (frontal axis) representing the pulleys. To simulate the pulley and spring action, internal joint forces were implemented to these revolute joints (Eq. 1).

$$M_{pulley}(Nm) = k_{spring}\left(\frac{N}{m}\right)\cdot r_{pulley}(m)^2\cdot\theta(rad) \qquad \text{Eq. 1}$$

The equivalent torque ($M_{pulley}$) depends on the spring constant ($k_{spring}$), the pulley radius ($r_{pulley}$) and the rotation of the pulley (Θ).

The pelvis belt was connected to the pelvis through a sagittal translation. A longitudinal axis revolute joint links the beam to the pelvis belt. This beam was modeled as 6 articulated rigid bodies, which is a simplified version of the slider-crank from the book *Symbolic Modeling of Multibody Systems* [20]. The internal articulated beam stiffnesses were implemented following to the articulated slider-crank's beam equation (Eq. 2 & Eq. 3).

$$Sagittal\ axis: K = \frac{E\cdot I}{d} = \frac{230Gpa*\{\frac{0,004m*(0,004m)^3}{12}\}}{0.1m} = 49.07Nm/rad$$

Eq.2

$$Frontal\ and\ transverse\ axis:$$

$$K = \frac{E\cdot I}{d} = \frac{230Gpa*\{\frac{0,01m*(0,004m)^3}{12}\}}{0.1m} = 1266{,}67Nm/rad$$

Eq.3

The stiffness K depends on the Young's modulus (E = 230GPa [26]) of the articulated beam, the geometrical moment of inertia (I) of a segment cross section (considering a segment had a rectangle area), relative to its geometrical center and the length of one out of the six segment (d).

Each body was linked by ball joints and a solid cut was used to close the loop between the upper trunk support and the pelvis belt. The upper trunk support was linked to the first thoracic vertebrae of the human model by a ball joint and was linked to the articulated beams by a longitudinal translation.

## 2.3. Experimental Setup

To validate the model, two lifting techniques were performed and compared with results from existing literature [15, 16, 17]. These lifting techniques were the stoop and the squat. The subject was asked to start in a standing position and to perform these movements while mimicking a lift (0 kg) and afterwards, while lifting a 15 kg box of dimensions 37cm x 25cm x 14.5

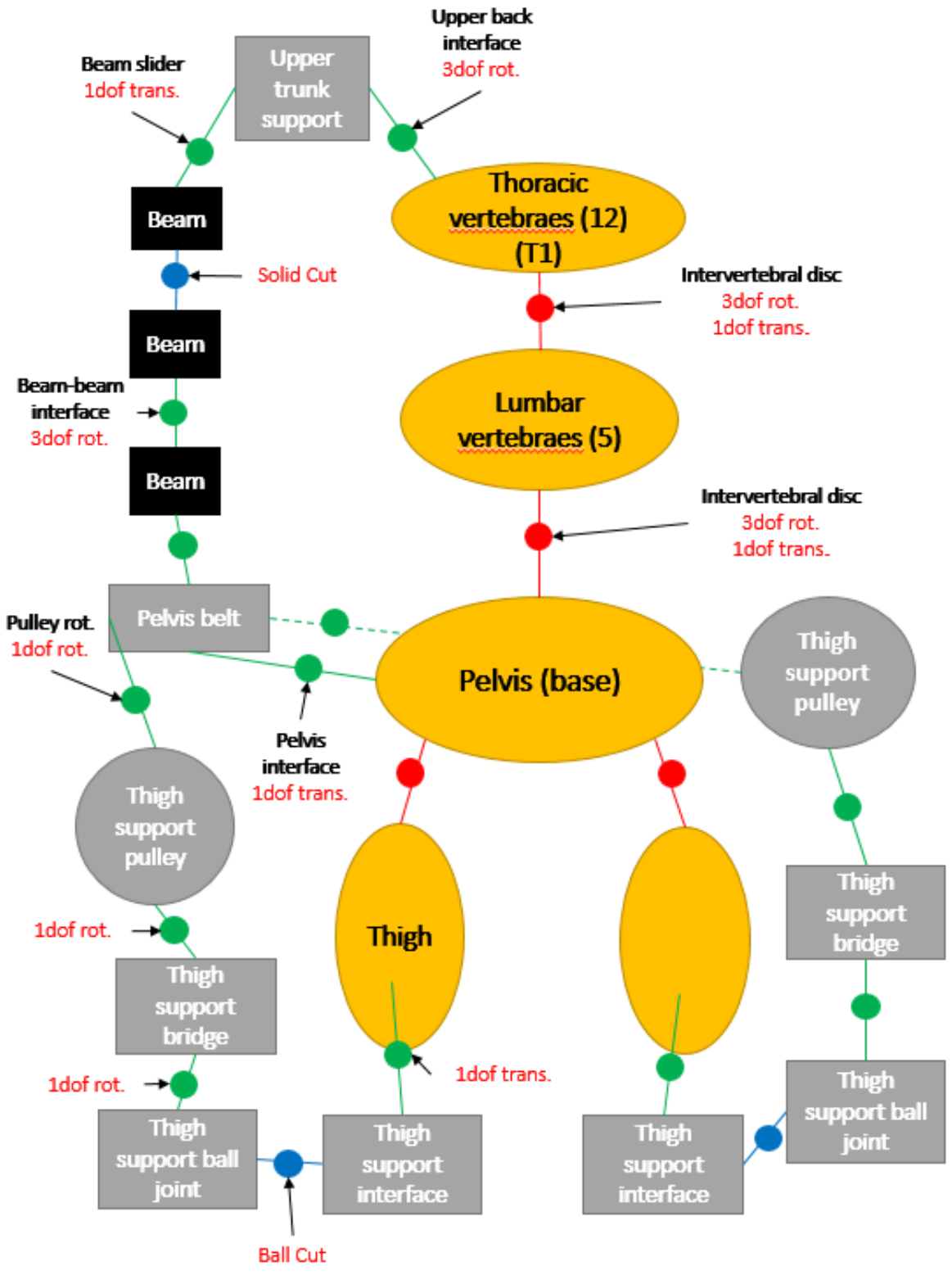

**Figure 3.** Exoskeleton linked to the human model

cm until subject reached back to initial standing position.

All movements were executed at subject pace, and without wearing the exoskeleton.

The acquisition system included 43 optokinetic sensors allowing to track the wrists elbows, shoulders, thoracic and lumbar spine, pelvis, hips, knee and ankle movements. A unique human model was created into the Vicon motion capture system. 12 motion-capture cameras (Vicon, UK) were used to capture 3D coordinates of the sensors with a sampling rate of 100Hz. The ground reaction applied to each foot was captured from AMTI force platforms with a sampling rate of 1000Hz. Only one cycle was recorded for every movement.

The data was then synchronized at 100Hz and sent to our MATLAB script to calculate range of motion or to assess the efforts in the joints following the data processing description (Section 2.4).

## 2.4. Data Processing

Globally, the joint torques and forces in the human body were computed by inverse dynamics, from the joint kinematic data, as well as the geometric and dynamic parameters. In details, two main steps were needed at every time iteration. The first step is a kinematic identification of every joint obtained with the body relative coordinates, values extracted from the anatomical landmarks captured using the Vicon system. The joints coordinates were obtained by processing the relative coordinates by inverse kinematics using ALGLIB Levenberg-Marquardt optimization algorithm [27]. The second step provided the efforts, by using the joint coordinates and through inverse dynamics based on dynamical equations system obtained from a Newton-Euler formalism (Eq. 4). ROBOTRAN generated the symbolic equations useful for the inverse dynamics [21]. The output variables were the efforts $\boldsymbol{Q}$, i.e. forces and torques, at each joint. They are dependent of the joint angles ($\boldsymbol{q}$), velocities ($\dot{\boldsymbol{q}}$), accelerations ($\ddot{\boldsymbol{q}}$), the external forces and torques ($\boldsymbol{frq}$, $\boldsymbol{trq}$) and the gravity ($\boldsymbol{g}$).

$$\boldsymbol{Q} = \phi(\boldsymbol{q}, \dot{\boldsymbol{q}}, \ddot{q}, \boldsymbol{frq}, \boldsymbol{trq}, \boldsymbol{g}) \qquad \text{Eq. 4}$$

# 3. Results

## 3.1. Joints Range of Motion and Lumbar Torque Validation Lumbar Torque

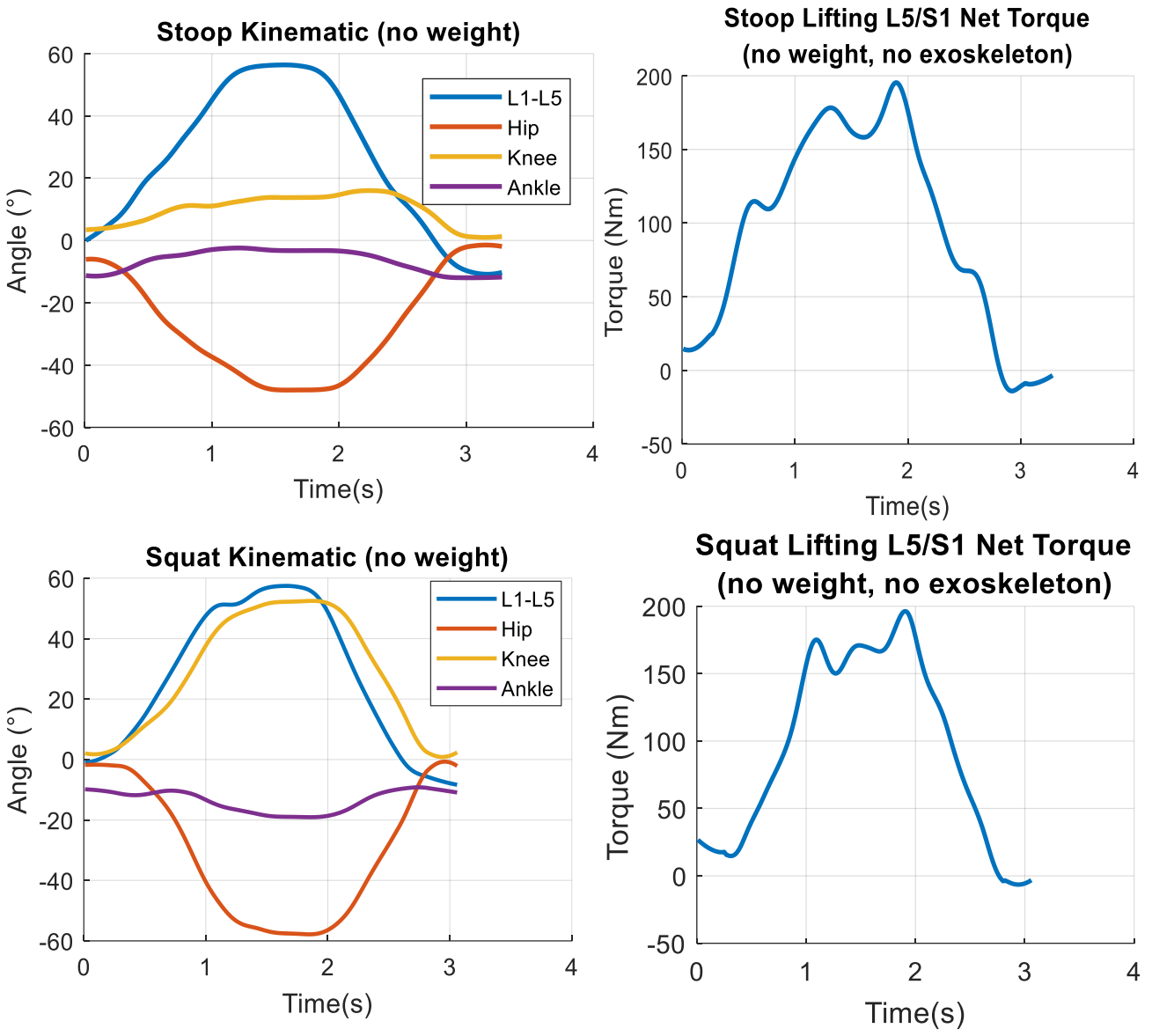

**Figure 4.** Stoop and squat joint's range of motion (left row) and L5/S1 net torque (right row)

The joint angles presented (Figure 4) are the angles from the sagittal plane. The angles for the hip, knee and ankle of each movement kinematic were extracted from the right side of the human model. Peak joint angles (absolute values) were found at 56.4°, 48.0°, 16.0° and 11.8° for the stoop while the peak values were found at 57.4°, 57.8°, 52.4° and 19.1° for the lumbar angle (i.e. sum of the angles from the L1 to L5 joints), hip, knee and ankle in the squat respectively. From the Figures 4 (left row), the major difference between the squat and the stoop can be observed at the knee joint. Right side of Figure 4 shows over time the net torque at the joint between the fifth lumbar vertebra and the sacrum (L5/S1). Peak torques are shown right after the lift-off motion and are 195.3 Nm for the stoop and 194.9 Nm for the squat.

## 3.2. Lifting Techniques Kinematics and Lumbar Torques with Weight

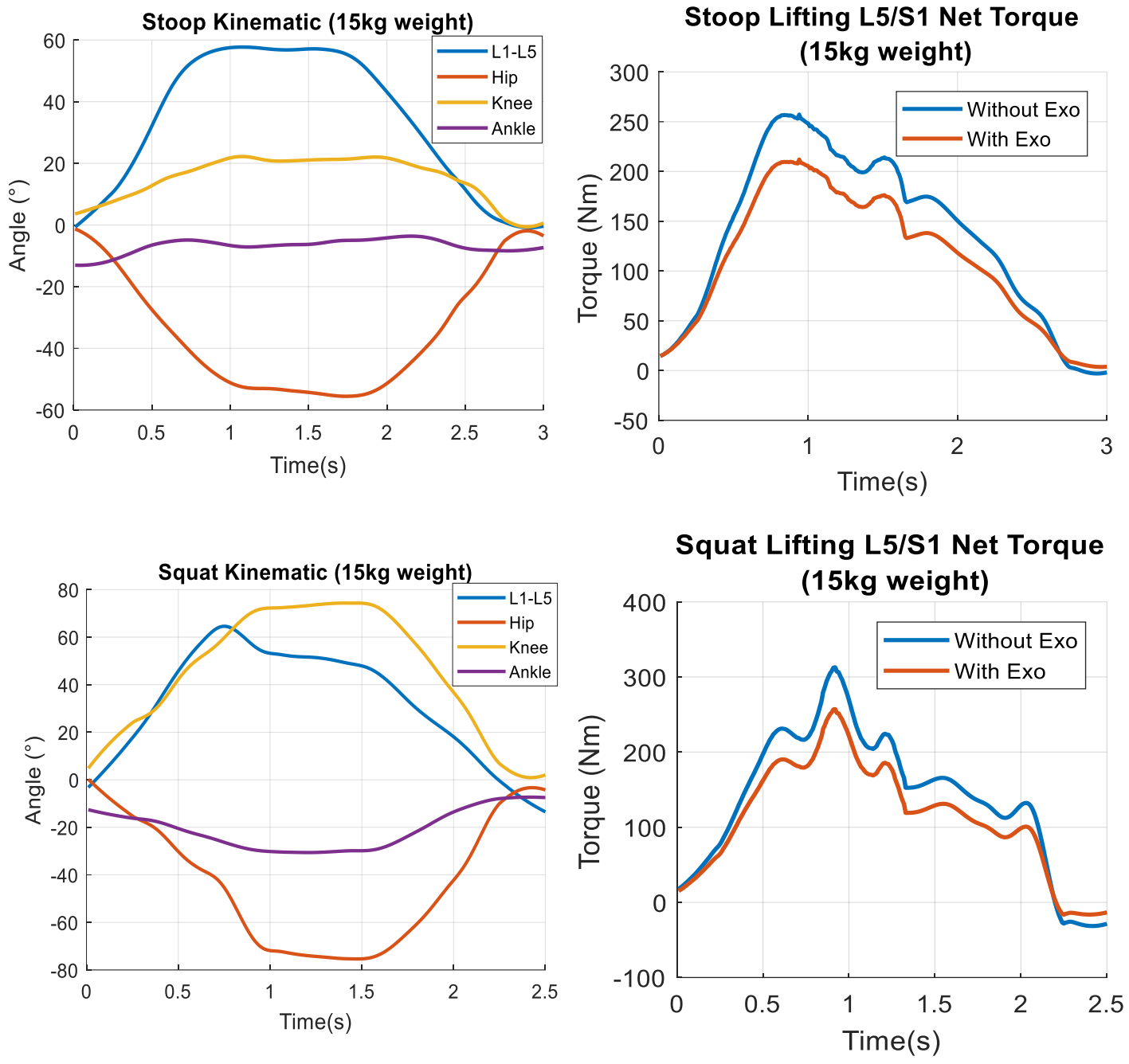

**Figure 5.** Stoop and squat joint's range of motion (left row) and L5/S1 net torque with and without the exoskeleton (right row)

Stoop joints kinematic in Figure 5 (upper left) is similar to Figure 4 stoop kinematic. Peak joint angles are 57.6°, 55.5°, 22.2° and 13.0° for the lumbar angle, hip, knee and ankle respectively. The time of the maximum amplitude appears to be longer and can be explained by the physical action of gripping and lifting the weight. The peak joint angles for the squat are 64.6°, 75.4°, 74.3° and 30.6° for the lumbar angle, hip, knee and ankle respectively. As for the lifting net torque measured without the assistance of the exoskeleton (Figure 5, right side), peak value for the stoop has been measured at 256.8 Nm, while the peak value for the squat was 312.9 Nm. As shown in Figure 5 (right side), the assistance of the exoskeleton reduced the torque to 209.8 Nm for the stoop and 257.2 Nm for the squat. Peak values can also be observed earlier in the lifting phase as the subject needs to generate enough torque to compensate the initial inertia of the box.

### 3.3. Human-Exoskeleton Interfacing Forces

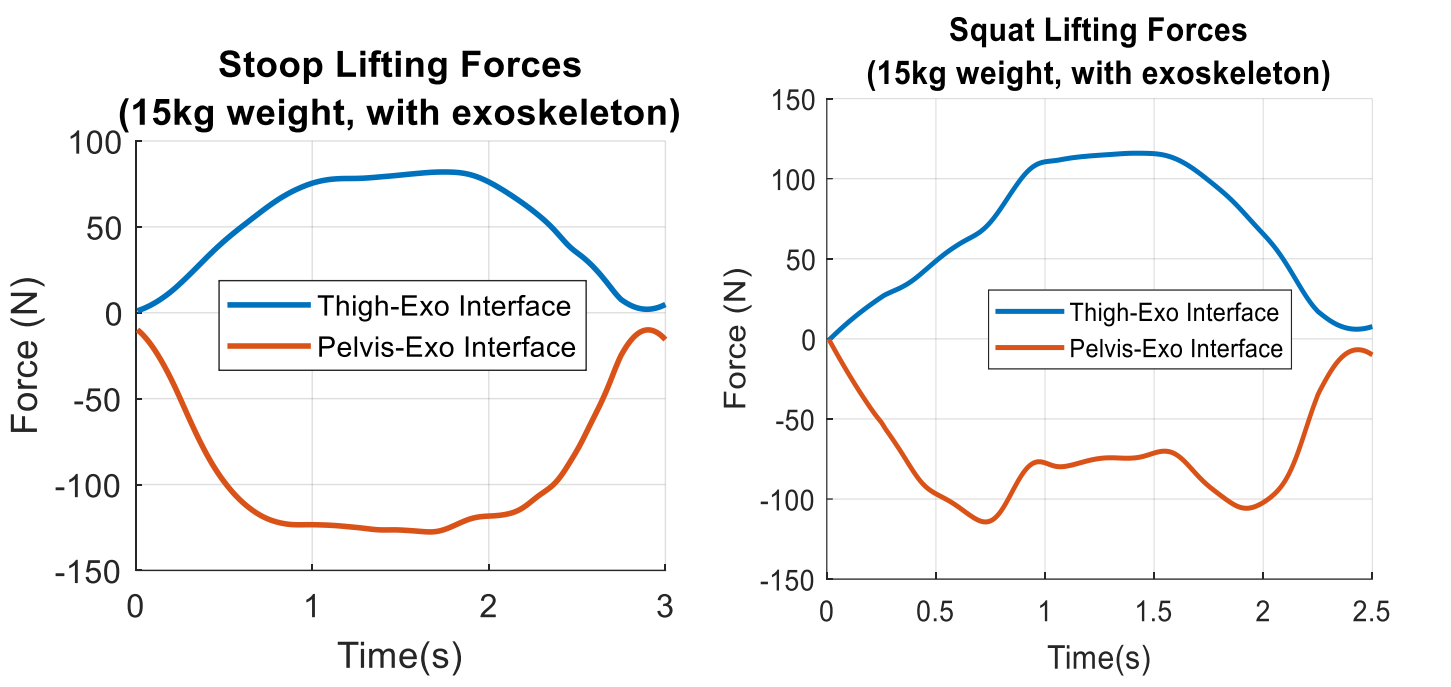

**Figure 6.** Interfacing normal forces between the right thigh and exoskeleton and the pelvis and the exoskeleton for the stoop (left) and the

Figure 6 shows the forces between the exoskeleton and the thighs and pelvis interfaces. For the stoop movement, maximum values (absolute) have been measured at 115.9 N for the thigh-exoskeleton interface and 114.3 N for the pelvis-exoskeleton interface. As for the squat, these interface forces were 81.9 N and 127.6 N for the thigh-exoskeleton and pelvis-exoskeleton interface, respectively.

## 4. Discussion

In this study, we have presented the interactions between a human body modeled with the thoracic and lumbar spine and an assistive trunk exoskeleton. The exoskeleton was designed by using composite articulated beams and tension springs. The presented model is the first one to simulate the lifting movement with a 3D human model by including the complete thoracic and lumbar spine. The model also takes into account the geometric convergence of the exoskeleton model, indicating that the device can actually complete its respective section movements without singularity.

The results from the inverse dynamic analysis showed similar results to the ones reported in the literature when the model was not equipped with the exoskeleton. Although the hip, knee and ankle joints and lumbar angle (i.e. L5/S1 in the literature) from the squat and the stoop without the weight (Figure 4, left side) showed similar results as the literature [15, 16], it seems that these values were different when compared to the same angles for each respective movement with a lifting weight of 15kg (Figure 5, left side). When the subject mimicked the lifting motion, peak torques were reported at 195.3 Nm for the stoop and 194.9 Nm for the squat. Torques obtained in the results were similar to the literature as Dolan & al. reported peak torques of 163±47Nm for the stoop and 190±64Nm for the squat [19].

As for the subject's movement with the added weight, peak torque for the stoop was 256.8 Nm, while the peak torque for the squat was 312.9 Nm. These values were similar to the literature. Dolan & al. reported peak torques of 273±79 Nm and 329±93 Nm for the squat with a 10kg and 20kg weight respectively while peak torques reported for the stoop were of 255±55 Nm and 280±96Nm for the 10kg and 20kg weights, respectively [19], while Manns & all reported a peak torque of 358.3 Nm for his OCP human model [6]. When equipped with the device, this study showed that peak values for the stoop was 209.8 Nm, while peak value for the squat was 257.2 Nm, meaning the exoskeleton was able to reduce the L5/S1 torque by 18.3% for the stoop and 17.7% for the squat when the subject lifted the 15 kg box.

The normal forces presented in section 3.3 seemed similar to the forces presented in Panero & al. optimized trunk exoskeleton results [7]. The interfacing forces at the back of the pelvis and on the thighs are explained by the need of the exoskeleton to produce opposite torque to the one in the L5/S1 joint to reduce the net L5/S1 joint net torque. The thighs contribute to this opposite torque, by being the contact point of the lever arm of the hip pulleys while the pelvis absorbs the torque generated by the articulated beam. This exoskeleton produces slightly less torque reduction than the state-of-the art passive lumbar support exoskeleton such as the PLAD, which provides torque reduction of 22-26% in the L4/L5 joint [5].

*Perspective*

The model composed of the human and exoskeleton could be further developed by:

- Optimizing the position of the articulated beam and the tension springs.
- Comparing the human model workspace with and without the exoskeleton equipped.
- Optimizing the tension springs values to reduce even more the L5/S1 net torque. This can be achieved using a similar method as presented by Blanchet & al [24].
- Assess the internal efforts of the articulated beam and the pulleys.
- Develop a model that included the muscle activity of the back alongside the internal forces of the back.
- Modeling a more realistic spring-pulley system to better assess the dynamic effects of this particular system.

## 5. Conclusion

The aim of this study was to present a novel 3D human dynamic multibody model equipped with a newly designed trunk assistance exoskeleton and to assess the effects of the device on the human body. The results showed that the exoskeleton diminished the net torque in the L5/S1 joint by 18.3% for the stoop and 17.7% for the squat. In order to reduce the torque, normal forces were applied at the exoskeleton's interface with the pelvis and the thighs. Finally, the novelty of this study is the modelling of the entire thoracic and lumbar spine.

## 6. Conflict of interest

The authors declare that they have no conflict of interest.

## 7. Acknowledgment

The authors would like to thank the Multibody research group of the Center for Research in Mechatronics at Université catholique de Louvain (UCL) for allowing the use of their software ROBOTRAN.